\documentstyle[12pt]{article}
\newcommand{\be}{\begin{eqnarray}}
\newcommand{\ee}{\end{eqnarray}}

\begin{document}
\thispagestyle{empty}

\vspace*{1cm}

\begin{center}
{\large{\bf Worldline Casting of the Stochastic Vacuum Model\\
and Non-Perturbative Properties of QCD:\\
General Formalism and Applications.\\}} \vspace{1.8cm}
{\large A.~I.~Karanikas and C.~N.~Ktorides}\\
\smallskip
{\it University of Athens, Physics Department\\
Nuclear \& Particle Physics Section\\
Panepistimiopolis, Ilissia GR 15771, Athens, Greece}\\
\vspace{1cm}
\end{center}
\vspace{0.4cm}

\begin{abstract}
The Stochastic Vacuum Model for QCD, proposed by Dosch and
Simonov, is fused with a Worldline casting of the underlying
theory, {\it i.e.} QCD. Important, non-perturbative features of
the model are studied. In particular, contributions associated
with the spin-field interaction are calculated and both the
validity of the loop equations and of the Bianchi identity are
explicitly demonstrated. As an application, a simulated
meson-meson scattering problem is studied in the Regge kinematical
regime. The process is modeled in terms of the ``helicoidal"
Wilson contour along the lines introduced by Janik and Peschanski
in a related study based on a AdS/CFT-type approach. Working
strictly in the framework of the Stochastic Vacuum Model and in a
semiclassical approximation scheme the Regge  behavior for the
scattering amplitude is demonstrated. Going beyond this
approximation, the contribution resulting from boundary
fluctuation of the Wilson loop contour is also estimated.
\end{abstract}

\newpage

\vspace*{.5cm}

{\bf 1. Introduction.}

\vspace*{.5cm}

The confrontation of non-perturbative issues associated with
dynamical processes constitutes a problem of great importance
which merits definite attention if QCD is to attain the status of
a complete and fully self-consistent theory. Clearly, the most
concrete advancement in formulating a non-perturbative casting of
QCD is traced to Wilson's proposal [1] which paved the way for the
lattice formulation of gauge field theories in general. Remarkable
results have been produced, especially in relation to the study of
static properties of hadrons [2], finite temperature properties of
the theory, etc.

On the analytical front, important theoretical progress, relevant
to non-perturbative, dynamical explorations of QCD, has been
achieved within the context of the loop equations [3-6], while, in
recent years, (super)string theory has, through the AdS/CFT
conjecture [7-9], opened new pathways for approaching the
non-perturbative domain of QCD, {\it albeit} in the sense of some
supersymmetric version of the theory and within the context of
unified schemes.

Generally speaking, the nontrivial aspects of QCD as a
relativistic gauge field theoretical system, stem from the fact
that the non-abelian gauge symmetry entering its description
incorporates an inherent non-linearity even before interaction
terms with matter field agents are introduced. Theoretical schemes
aiming at a heads-on analytical confrontation of non-linear
quantum field systems do, of course, exist, possibly the most
concrete one being expressed in terms of the infinite battery of
the Schwinger-Dyson equations. Even in this case however, the
relevant computational procedure for the solution of these
(integral) equations is not only methodologically complex but,
more importantly, unless totally summed, there is no {\it a
priori} guarantee that they are in position to capture the full
non-perturbative content of any given field theoretical system
and/or describe its expected various phases.

A notable field theoretical approach aiming at the study of
non-perturbative issues in QCD, such as confinement, chiral
symmetry breaking {\it etc,} has been proposed by Dosch and
Simonov [10-12] and goes by the name of Stochastic Vacuum Model
(SVM). By design, the construction of the model takes into
consideration the non-trivial structure of the QCD vacuum state
[13], while, at the same time, it secures a role for the Stokes'
theorem (non-abelian casting thereof) through which
electric-magnetic duality issues can be fully taken into account.
The basic building blocks of the SVM scheme are the, so called,
{\it field strength correlators}, the definition of which will be
given later, while for its computational strategy it employs the
so-called Field (strength) Correlators Method, FCM for short. For
the reader not familiar with the SVM, we hope that the information
provided in this paper will sufficiently illustrate the reasoning
behind its definition, as well as  its properties and physical
content. For a deeper insight to the model, one is referred to the
original papers [10-12] and/or the excellent review articles
[14,15] which also present a variety of its applications.

The characteristic aspect of the present work is that it adopts,
as basic methodological tool, the Worldline casting of gauge field
theoretical systems [16] appropriately adjusted to the SVM.  Our
goal is to confront genuinely non-perturbative issues associated
with dynamical processes of physical interest. At the same time
-and at a purely theoretical level- we hope that the present
effort will offer new insights further promoting the effectiveness
of the SVM
 as a credible and viable theoretical tool for exploring
the non-perturbative content of QCD.

The aim of the first part of this paper (Sections 2 and 3) is to
accomplish the task of carrying out cumbersome computations which
reveal fundamental properties of the SVM that are of immediate
relevance to our specific purposes and to establish the
consistency of the model with the loop equations [3-6], as well as
the Bianchi identity for QCD. Clearly, such an occurrence
strengthens the credibility of the SVM as a theoretical
construction which is consistent with QCD as a whole, {\it i.e.}
in the sense that it goes beyond perturbation theory. Our main
theoretical application will be realized in the second part
(Sections 4 and 5)where we undertake the description of a,
theoretically simulated, meson-meson scattering process at the
high energy, small momentum transfer kinematical (Regge) regime.
The relevant description of such a dynamical process necessarily
``protrudes" into the non-perturbative domain of QCD as
represented, in our case, by the SVM.

 Our presentation is organized as follows. In section (2), we
 discuss general aspects which prepare the ``fusing" of the Worldline description of a non-abelian gauge
 field theoretical system, such as QCD, with the SVM. We start by focusing our attention
 on a situation where a matter field entity of spin
 $j$ interacts with a set of non-abelian gauge field modes as it
 propagates along a closed contour in (Euclidean) space-time, hence
 subjected to a spin-field interaction. The basic dynamical content of the process will be displayed by two alternative
 formulations. The first focuses on the (closed) Wilson
 contours traced by the particle entity. The second is based on a shifted field strength tensor that is integrated
 over an arbitrary surface bounded by the closed contour. Obviously the (non-abelian)
 Stokes' theorem plays a central role in relating the two
 descriptions, an occurrence of central importance to our purposes,
 given that the Stokes' theorem enters the SVM scheme in a major way.

We shall subsequently introduce the so-called cluster expansion
[14, 15, 17], which employs the field strength correlators, the
basic dynamical quantities of the overall description. In fact,
the cluster expansion is the essence of the stochastic nature of
the model and provides the key element for quantifying the
stochastic vacuum hypothesis. Once the aforementioned task is
accomplished, we shall be in position to derive, as a first
general result, the equation that determines the surface over
which the two-point correlator must be integrated.

 Section 2 deals also with fairly demanding calculations
within the context of the Worldline formalism whose first result
is the derivation of an explicit expression for the spin factor.
The latter represents the genuinely non-perturbative, spin-field
interaction dynamics and necessarily enters [18] the analysis of
the meson-meson scattering process. Section 3 is devoted to the
verification of the loop equations and of the Bianchi identity in
the framework of the adopted approach. We consider these results
to be of importance since they solidify the credibility of the SVM
as a construction which reproduces sound, theoretical properties
of QCD and demonstrate the compatibility with its non-perturbative
content.

In Section 4 we present a semiclassical calculation of a simulated
meson-meson scattering amplitude. The Wilson loop that carries the
dynamics of the process is a helicoidal embedded in a
4-dimensional background. The calculation is performed in the
framework of the SVM and lead us to a Regge-type behavior for the
amplitude valid in the physical region of the scattering.
Corrections related to the fluctuations of the aforementioned
Wilson contour will be discussed in Section 5. In the same Section
the contribution of the spin-factor is examined. Some concluding
comments will be made in the closing Section.

 \vspace*{.5cm}

 {\bf 2. Worldline Formalism and Field Strength Correlators.}

\vspace*{.5cm}

In this section, certain basic features of the Field Correlator
Method [15] will be reviewed in the context of the Worldline
casting of a Quantum, Gauge Field Theoretical System in
interaction with a matter particle mode of a given spin $j$. More
explicitly, the particle entity is taken to propagate along a
given, closed, contour (worldline) while interacting with a
dynamical set of non-abelian gauge fields ${\cal A}$. According to
our introductory discussion, the main objective is to introduce
the necessary tools to facilitate non-perturbative, theoretical
explorations of QCD in the framework of the SVM.

 One of the most important advantages of the
 Worldline formalism is that it allows one to reduce the
physical amplitudes to weighted integrals of averaged Wilson loops
[16]. In this connection, one can apply powerful techniques, such
as the cluster expansion, both at the perturbative (in the sense
of series resummation) and at the non-perturbative level. For the
latter, the Worldline formalism proves to be quite crucial because
one can develop methods based on background gauge fixing strategy
[18-22], which enables one to treat the non-perturbative fields as
background.

Let us, then, consider a particle entity of spin $j$ propagating
from some point and back to the same point while interacting
dynamically with a non-abelian gauge field system  ${\cal A}$. The
basic structure of the quantum mechanical amplitude associated
with such a process is written in the Worldline formalism (all
indices suppressed; Euclidean formalism adopted) as follows [16]
\begin{equation}
 K(L)=Tr\int\limits_{x(0)=x(1)}{\cal
D}x(\tau)\exp\left(-{1\over4L}\int_0^1
d\tau\dot{x}^2\right)\left\langle P\exp\left(i\int_0^1
d\tau\dot{x}\cdot {\cal A}+L\int_0^1 d\tau J\cdot
F\right)\right\rangle_{\cal A},
\end{equation}
where it should be noted that the parameter $L$ has dimensions
$m^{-2}$ and must be integrated over through a weight factor
$\exp(-Lm^2)$ in order to obtain a result with physical content.
The matrices $J_{\mu\nu}$ stand for the Lorentz generators,
pertaining to the spin of the propagating entity. Accordingly, the
last term represents the spin-field interaction.

The above expression for $K(L)$ can be recast into the form [16]
\begin{equation}
 K(L)=Tr\int\limits_{x(0)=x(1)}{\cal
D}x(\tau)\exp\left(-{1\over4L}\int_0^1 d\tau\dot{x}^2\right)
P\exp\left({i\over2}L\int_0^1 d\tau
J\cdot\frac{\delta}{\delta\sigma}\right)
 \left\langle P\exp\left(i\int_0^1 d\tau\,\dot{x}\cdot {\cal A}\right)\right\rangle_{\cal A},
 \end{equation}
 where
 \begin{equation}
 \frac{\delta}{\delta\sigma_{\mu\nu}(x(\tau))}=\lim\limits_{\eta\to
 0}\int\limits_{-\eta}^{\eta}dh\,h\frac{\delta^2}{\delta x_\mu\left(\tau+{h\over2}\right)
  \delta x_\nu\left(\tau-{h\over2}\right)}
 \end{equation}
defines a regularized expression for the area derivative [3-6].

Strictly speaking, expression (2) has a well defined meaning only
for smooth [23] loops. On the other hand, when such expressions
are used for the purpose of describing physically interesting
processes the contour is forced to pass through points $x_i$ where
momentum is imparted by an external agent (field). Such a
situation is mathematically realized by inserting a corresponding
chain of delta functions $\delta[x(\tau_i)-x]$ in the integral,
which produces a loop with cusps, in which case the action of the
area derivative operator entering (3) must be understood
piecewise, {\it i.e.},
\begin{equation}
P\exp\left({iL\over2}\int\limits_0^1d\tau
J\cdot\frac{\delta}{\delta\sigma}\right) =\cdot\cdot\cdot
P\exp\left({iL\over2}\int\limits_{\tau_1}^{\tau_2}d\tau
J\cdot\frac{\delta}{\delta \sigma}\right)
P\exp\left({iL\over2}\int\limits_0^{\tau_1}d\tau
J\cdot\frac{\delta}{\delta \sigma}\right).
\end{equation}

The first step towards the application of the FCM is taken by
employing the non-abelian Stokes' theorem [24] with the help of
which one can write (the symbol $P_s$ stands for surface ordering
[14])
\begin{equation}
W[C]\equiv{1\over N_C} Tr\left\langle P\exp\left(i\oint_Cdx\cdot
{\cal A}\right)\right\rangle_{\cal A}={1\over N_C}Tr\left\langle
P_s\exp\left[i\int\limits_{S(C)}dS_{\mu\nu}(z)G_{\mu\nu}(z,x_0)\right]\right\rangle_{\cal
A},
\end{equation}

The above expression is valid for any loop $C$ with disc topology,
irrespectively of the surface $S(C)$. We also mention that for the
area element we adopt the standard expression
\begin{equation}
dS_{\mu\nu}={1\over2}d^2\xi\sqrt{g}t_{\mu\nu}(\xi),\quad
t_{\mu\nu}(\xi)={1\over\sqrt{g}}\epsilon^{ab}\partial_a z_\mu
\partial_b z_\nu,\quad a,   b=1,2;\,(\xi^1,\xi^2)=(\tau,s).
\end{equation}
Finally, in relation (5) we have set [14,15]
\begin{equation}
G_{\mu\nu}(x_0,z)=\phi(x_0,z)F_{\mu\nu}(z)\phi(z,x_0)
\end{equation}
with
\begin{equation}
\phi(z,x_0)=P\exp\left(i\int\limits_{x_0}^z dw\cdot {\cal
A}\right)
\end{equation}
a phase factor [14,15] which is a parallel transporter known also,
in the SVM nomenclature [17], as {\it connector}. The reference
point $x_0$ is chosen arbitrarily on the surface $S$; arbitrary is
also the curve that joins the points $x_0$ and $z$.

It can be proved [24] that (5) depends neither on the surface nor
on the contour used to define the connector (8), as long as the
non-abelian Bianchi identities are satisfied. In the loop language
such a requirement can be cast into the following statement:

The relation
\begin{equation}
\frac{\delta}{\delta z_\lambda(\xi)}Tr\left\langle
P\exp\left[i\int\limits_{S(C)}dS_{\mu\nu}(z)G_{\mu\nu}(z,x_0)\right]\right\rangle_{\cal
A}=0
\end{equation}
is valid independently of the surface choice provided that
\begin{equation}
\epsilon^{\kappa\lambda\mu\nu}\partial_\lambda^{x(\tau)}\frac{\delta}{\delta\sigma_{\mu\nu}(x(\tau))}W[C]=0,
\end{equation}
which corresponds to the Bianchi identity for the gauge system
[6].

The simplest way to prove the non-abelian Stokes theorem is to
adopt the contour gauge [15]
\begin{equation}
{\cal A}_\mu(x)=\int_0^1ds\,\partial_s z_\kappa(s,x)\frac{\partial
}{\partial x_\mu}z_\lambda(s,x)F_{\kappa\lambda}(z(s,x)),
\end{equation}
with $\{z_\mu(s,x),\,s\in[0,1]\}$ an arbitrary, smooth curve from
the reference point $x_0$ to some point $x$:
\begin{equation}
z_\mu(0,x)=x_{0\mu},\quad z_\mu(1,x)=x_\mu.
\end{equation}
Indeed, using Eq. (11) one can immediately see that
 \begin{equation}
 \oint dx_\mu {\cal A}_\mu(x)=\int_0^1d\tau\int_0^1ds\partial_s
 z_\mu\partial_\tau
 z_\nu F_{\mu\nu}(z)={1\over2}\int_0^1d\tau\int_0^1ds\epsilon^{ab}\partial_az_\mu\partial_bz_\nu
 F_{\mu\nu}(z)\\
 ={1\over2}\int d^2\xi\sqrt{g}t_{\mu\nu}(z)F_{\mu\nu}(z).
\end{equation}
For the gauge choice (11) the vector potential satisfies the
condition $t_\mu(x){\cal A}_\mu(x)=0$, with
$t_\mu(x)=\partial_sz(s,x)|_{s=1}$, which implies that the
connector (8) can be considered  as the unit matrix in the contour
gauge. In any case, the presence of the connectors in (7)
guarantees gauge invariance.

The next step towards the application of the FCM is to introduce
the so-called cluster expansion [10-12,14,15,17] for the Wilson
loop, formally written as
\begin{equation}
W[C]={1\over
N_c}Tr\exp\left(\sum\limits_{n=1}^\infty\frac{i^n}{n!}\int\limits_{S(C)}
   dS_{\mu_n\nu_n} \cdot\cdot\cdot
dS_{\mu_1\nu_1}<<G_{\mu_n\nu_n}(z_n,x_0)\cdot\cdot\cdot
G_{\mu_1\nu_1}(z_1,x_0)>>\right),
\end{equation}
where the symbol $<<\cdot\cdot\cdot>>$  translates as follows:
\begin{equation}
\begin{array}{lllll}
<<O(1)>>=<O(1)>\\
<<O(1)O(2)>>=<P_s(O(1)O(2))>-{1\over2}<(O(1)><O(2)>-{1\over2}<(O(2)><O(1)>\\
<<O(1)O(2)O(3)>>=<P_s \left(O(1)O(2)O(3)\right)>-{1\over2}(<P_s
(O(1)O(2))><O(3)>+{\rm cycl. perm.})\\
\quad\quad +{1\over3}(<O(1)><O(2)><O(3)>+{\rm cycl. perm.})\\
+\cdot\cdot\cdot.
\end{array}
\end{equation}
and is reminiscent of the cluster expansion in Statistical
Mechanics.
 It is pointed out that, due to the color
 neutrality of the vacuum, expectation values of all correlators in
 (14) are proportional to the unit matrix in color space. This
 makes color ordering unnecessary.

 The formula of Eq.(14) quantifies the formulation of the SVM. It turns out [14,15]that the most
important contribution to
 the cluster expansion comes from the two-point correlator:
\begin{equation}
\Delta^{(2)}_{\mu\nu,\lambda\rho}(z-z')={1\over N_c}Tr\langle
G_{\mu\nu}(z,x_0)G_{\lambda\rho}(z',x_0)\rangle_A={1\over
N_c}Tr\langle
F_{\mu\nu}(z)\phi(z,z')F_{\lambda\rho}(z')\phi(z',z)\rangle_A.
\end{equation}
The above defines the {\it field strength correlator}, a quantity
which constitutes the basic building block of the model.

 Some natural assumptions are incorporated in the above
definition. The first has to do with the Lorentz invariance of the
vacuum, which is explicitly indicated in the left hand side of
(16) by the fact that the correlator depends on the {\it distance}
between the points $z$ and $z'$. The second is that, on the other
hand, the correlator does not depend on (the gauge parameter)
$x_0$. This is a credible assumption, taking into account the fact
that we have been working, from the very beginning, with a gauge
invariant amplitude. Accordingly, the basic assumption of the SVM
leads to the statement that
\begin{equation}
W[C]\propto\exp\left[-{1\over2}\int\limits_{S(C)}dS_{\mu\nu}(z)\int\limits_{S(C)}dS_{\lambda\rho}(z')
\Delta^{(2)}_{\mu\nu,\lambda\rho}(z-z')\right],
\end{equation}
where the surface element enters through the use of the
(non-abelian) Stokes' Theorem.

Two important points should now be made. First, the last
expression is supposed to be valid in a certain limit. Explicitly,
it is assumed [10-12,14,15] that the vacuum fluctuations establish
a {\it correlation length} $T_g$ beyond which correlations decay
very fast. If $\Delta$ is an order of magnitude estimation for the
two point correlator, relation (17) is considered as an asymptotic
approximation which is valid in the limit $T^2_g\sqrt{\Delta}\to
0$. The second is that, while expression (14) does not depend on
the particular surface one uses for the application of the
non-abelian Stokes' theorem, approximation (17) does. Thus, the
stochasticity assumption transforms relation (9) to an equation
which determines the dominant surface in the cluster expansion. To
quantify this statement we write
\begin{equation}
A[C]={1\over2}\int\limits_{S(C)}dS_{\mu\nu}(z)\int\limits_{S(C)}dS_{\lambda\rho}(z')
\Delta^{(2)}_{\mu\nu,\lambda\rho}(z-z'),
\end{equation}
from which it is easily determined that
\begin{equation}
\frac{\delta A}{\delta z_\sigma
(\tilde{\xi})}=-{1\over2}\sqrt{g(\tilde{z})}t_{\mu\nu}(\tilde{z})
\int\limits_{S(C)}dS_{\lambda\rho}(z')\left[\tilde{\partial}_\mu
\Delta^{(2)}_{\sigma\nu,\lambda\rho}(\tilde{z}-z')
+\tilde{\partial}_\nu\Delta^{(2)}_{\mu\sigma\,\lambda\rho}
(\tilde{z}-z')+\tilde{\partial}_\sigma\Delta^{(2)}_{\nu\mu,\lambda\rho}(\tilde{z}-z')\right],
\end{equation}
where we have written $\tilde{\partial}_\mu\equiv
\frac{\partial}{\partial\tilde{z}_\mu}$. The calculation of the
derivative of the correlator (16) needs to take into account [15]
the presence of the connectors $\phi(z,x_0)$. In Appendix A we
show that
\begin{equation}
\frac{\partial}{\partial z_\mu}\phi(z,x_0)=i{\cal
A}_\mu(z)\phi(z,x_0)-iI_\mu(z,x_0),
\end{equation}
with
\begin{equation}
\begin{array}{cc}
I_\mu(z,x_0)=\int\limits_0^1dt\,(\partial_t\omega_\kappa)\,\frac{\partial\omega_\lambda}{\partial
z_\mu}\phi (z,\omega)
F_{\kappa\lambda}(\omega)\phi(\omega,x_0)\\
\omega=\omega(t,z),\,\omega(1,z)=z, \quad\omega(0,z)=x_0.
\end{array}
\end{equation}
Given the above, we conclude that
\begin{equation}
\partial_\mu G_{\alpha\beta}(z,x_0)=\phi(x_0,z)D_\mu
F_{\alpha\beta}(z)\phi(z,x_0)+i[I_\mu(z,x_0),G_{\alpha\beta}(z,x_0)]
\end{equation}
and consequently write
\begin{equation}
\partial_\mu \tilde{G}_{\mu\nu}(z,x_0)=\phi(x_0,z)D_\mu
\tilde{F}_{\mu\nu}(z)\phi(z,x_0)+i[I_\mu(z,x_0),\tilde{G}_{\mu\nu}(z,x_0)],
\end{equation}
where we have set
$\tilde{G}_{\mu\nu}={1\over2}\epsilon^{\mu\nu\alpha\beta}G_{\alpha\beta}$.

The above analysis establishes the following relation [15] among
the derivatives of the correlator:
\begin{equation}
{1\over2}\epsilon^{\sigma\kappa\mu\nu}\partial_\sigma\Delta^{(2)}_{\mu\nu,\lambda\rho}
\equiv\partial_\sigma\tilde{\Delta}^{(2)}_{\sigma\kappa,\lambda\rho}
={1\over N_c}Tr\langle
D_\sigma\tilde{F}_{\sigma\kappa}(z)F_{\lambda\rho}(z')\phi(z',z)\rangle_{\cal
A}-\Delta_{\kappa\lambda\rho}(z,z'),
\end{equation}
where
\begin{equation}
\Delta_{\kappa\lambda\rho}(z,z')={1\over
N_c}Tr\langle\tilde{F}_{\sigma\kappa}(z)I_\sigma(z,z')F_{\lambda\rho}(z')\phi(z,z')-\tilde{F}_{\sigma\kappa}(z)
\phi(z,z') F_{\lambda\rho}(z')I_\sigma(z',z)\rangle_{\cal A}.
\end{equation}
Thus, if one assumes the validity of the Bianchi identities
$D_\mu\tilde{F}_{\mu\nu}=0$, one concludes that
\begin{equation}
\partial_\mu\Delta^{(2)}_{\sigma\nu,\lambda\rho}+
\partial_\nu\Delta^{(2)}_{\mu\sigma,\lambda\rho}
+\partial_\sigma\Delta^{(2)}_{\nu\mu,\lambda\rho}=\epsilon^{\mu\kappa\sigma\nu}\Delta_{\kappa\lambda\rho}.
\end{equation}
Accordingly, Eq (9) can be represented as follows
\begin{equation}
\frac{\delta A}{\delta
z_\sigma(\tilde{\xi})}={1\over2}\sqrt{g(\tilde{z})}t_{\mu\nu}\int\limits_{S(C)}dS_{\lambda\rho}(z)
\epsilon^{\sigma\kappa\mu\nu}\Delta_{\kappa\lambda\rho}(\tilde{z}-z)=0.
\end{equation}
It is worth noting that $\Delta_{\kappa\lambda\rho}$ is a three
point correlation function and it would be {\it identically zero}
if we were considering an Abelian gauge theory so that the above
relation becomes, really, an identity, telling nothing about the
particular surface involved in Stokes' theorem, an expected result
given that relation (17) is exact in the framework of QED. An
extensive discussion of the physical content of the correlator
$\Delta_{\kappa\lambda\rho}$ can be found in [14,15]. According to
the analysis presented in the aforementioned references
confinement in QCD  occurs due to the non-zero value of the
(non-abelian) correlator $\Delta_{\kappa\lambda\rho}$. Eq.(27)
indicates that this correlator also defines the relevant surface
on which the two-point correlator ``lives".

Now, it has been demonstrated [10-12,14,15] that, in the
asymptotic limit $\mid\tilde{z}-z\mid\gg T_g$, Eq.(27) determines
the surface $S$ bounded by the contour $C$ as the minimal one. To
demonstrate this we employ the following general Lorentz structure
representation [14,15] for the two-point correlation function:
\begin{equation}
\Delta^{(2)}_{\mu\nu,\lambda\rho}(\bar{z})=(\delta_{\mu\lambda}\delta_{\nu\rho}-\delta_{\mu\rho}\delta_{\nu\lambda})
D(\bar{z})+\left \{ {1\over2}\frac{\partial}{\partial\bar{z}_\mu}
[(\bar{z}_\lambda\delta_{\nu\rho}
-\bar{z}_\rho\delta_{\nu\lambda})D_1(\bar{z})]-(\mu\leftrightarrow\nu)\right\},
\end{equation}
where we have set $\bar{z}=z-z'$.

It is easy to see that
\begin{equation}
{1\over2}\epsilon^{\mu\kappa\sigma\nu}\partial_\sigma\Delta^{(2)}_{\mu\nu,\lambda\rho}
=\epsilon^{\kappa\lambda\rho\sigma}\partial_\sigma D=
\Delta_{\kappa\lambda\rho}.
\end{equation}
With the help of the above relation Eq.(27) can be cast into the
form
\begin{equation}
\frac{\delta}{\delta
z_\sigma(\tilde{\xi})}\int\limits_{S(C)}dS_{\mu\nu}(z)\int\limits_{S(C)}dS_{\mu\nu}(z')D(z-z')=0.
\end{equation}

The functions $D$ and $D_1$ have been measured in lattice
calculations [25] and have been found to be of very fast decrease
as $\mid z-z'\mid^2\to \infty$. In the considered region both of
them were found to be of the form $f\left(\frac{\mid
z-z'\mid^2}{T^2_g}\right)$ and that they go exponentially fast to
zero for $\mid z-z'\mid>T_g$. In this region, we write:
\begin{equation}
z(\xi')\simeq z(\xi)+(\xi'-\xi)^a\partial_a z(\xi),\quad \mid
z-z'\mid^2\simeq (\xi'-\xi)^a (\xi'-\xi)^b g_{ab}.
\end{equation}

In the considered limit and taking into account that $t_{\mu\nu}
t_{\mu\nu}=2$ we find
\begin{equation}
\frac{\delta}{\delta z_\sigma(\tilde{\xi})}\left[\sigma S+{\cal
O}(T_g^4\Delta)\right]=0,
\end{equation}
 where the string tension
\begin{equation}
 \sigma\equiv\frac{T^2_g}{2}\int d^2w D(w^2),
 \end{equation}
has been introduced and where we have also written
\begin{equation}
S\equiv\int d^2w\sqrt{g(w)}
\end{equation}
for the area of the surface bounded by the Wilson curve. In the
last equations we used the dimensionless parameter
$w_\mu=\frac{1}{T_g}(z-z')_\mu$ and we have written
$g_{ab}(w)=\partial_aw_\mu\partial_bw_\mu$ the induced metric. It
accordingly follows that in the limit $T_g\to 0$, the surface on
which the two-point correlation dominates is the minimal one.

Now we turn our attention on the {\it spin factor}, whose role is
to incorporate the spin-field interaction in the framework of the
worldline formalism. We mention that in the present work we shall
be dealing with massive fermions. It should be noted, at the same
time, that the spin factor can also be extended [21,22] for the
case of, massless, bosons of spin-1. Accordingly, the calculations
to be presented in this section can easily be extended to bosonic
fields.

We start by inserting into the SVM formula for the Wilson loop,
{\it cf.} Eq.(17), into the worldline integral expression (1), the
basic goal being that of calculating the spin-field interaction
with the help of the area derivative operator defined in (3). Once
this is accomplished the stage will be in place for performing
specific calculations of physical interest.

We start by introducing the spin factor by formally casting Eq.(1)
into the form
\begin{equation}
 K(L)=Tr\int\limits_{x(0)=x(1)}{\cal
D}x(\tau)\exp\left(-{1\over4L}\int_0^1
d\tau\dot{x}^2\right)\Phi^{(j)}[C]e^{-A[C]},
\end{equation}
where the quantity
\begin{equation}
\Phi^{(j)}[C]\equiv e^{A[C]}P\exp\left({i\over
2}\int_0^1d\tau\,J\cdot\frac{\delta}{\delta\sigma}\right)e^{-A[C]}
\end{equation}
defines the spin factor characterizing a particle entity
propagating on the Wilson curve  $A[C]$. Its calculation is not
trivial and requires a number of steps the first of which is to
determine the action of the area derivative operator on $A$.

The first objective of the computation of the spin factor is to
study the change on $A[C]$ induced by an infinitesimal variation
of the boundary. The relevant problem is formulated as follows
\begin{eqnarray}
\frac{\delta A}{\delta
x_\mu(\tau_1)}&=&\int\limits_0^1ds[a(\tau_1,s)\dot{z}_\alpha(\tau_1,s)]'\int\limits_{S(C)}dS_{\gamma\delta}(z')
\Delta^{(2)}_{\alpha\mu,\gamma\delta}\nonumber +\\
&+&\int\limits_0^1ds\,
a(\tau_1,s)\dot{z}_\alpha(\tau_1,s)z'_\beta(\tau_1,s)\int\limits_{S(C)}dS_{\gamma\delta}(z')
\left(\partial_\mu\Delta^{(2)}_{\alpha\beta,\gamma\delta}+\partial_\alpha\Delta^{(2)}_{\beta\mu,\gamma\delta}
\right),
\end{eqnarray}
where the dot denotes (partial) derivation with respect to $\tau$,
while the prime derivation with respect to $s$. Moreover,
 the correlators depend on the distance
$\mid z(\tau_1,s)-z(\tau',s')\mid$ and we have written
$\frac{\delta z_\alpha(\xi)}{\delta
x_\mu(\tau_1)}=\delta_{\alpha\mu}\delta(\tau-\tau_1)\alpha(\tau_1,s)$.

Using Eq. (26)  equation (37) is recast into the form
\begin{eqnarray}
\frac{\delta A}{\delta
x_\mu(\tau_1)}&=&\int\limits_0^1ds[a(\tau_1,s)\dot{z}_\alpha(\tau_1,s)]'
\int\limits_{S(C)}
dS_{\gamma\delta}(z')\Delta^{(2)}_{\mu\alpha,\gamma\delta}-\nonumber\\
&-&\int\limits_0^1ds\,a(\tau_1,s)\dot{z}_\alpha(\tau_1,s)z'_\beta(\tau_1,s)\int\limits_{S(C)}
dS_{\gamma\delta}(z')(\partial_\beta\Delta_{\mu\alpha,\gamma\delta}^{(2)}-\epsilon^{\mu\nu\alpha\beta}
\Delta_{\gamma\nu\delta})=\nonumber\\
&=&\dot{x}_\alpha(\tau_1)\int\limits_{S(C)}dS_{\gamma\delta}(z')\Delta_{\alpha\mu,\gamma\delta}^{(2)}
[x(\tau_1)-z(\tau',s')]+\nonumber\\
&+&\int\limits_0^1ds\,a(\tau_1,s)\dot{z}_\alpha(\tau_1,s)z'_\beta(\tau_1,s)
\int\limits_{S(C)}dS_{\gamma\delta}(z')\epsilon^{\mu\nu\alpha\beta}
\Delta_{\nu\gamma\delta}[z(\tau_1,s)-z(\tau',s')].
\end{eqnarray}
The last term in the above relation is zero on account of
condition (27):
\begin{eqnarray}
&&\dot{z}_\alpha(\tau_1,s)z'_\beta(\tau_1,s)
\int\limits_{S(C)}dS_{\gamma\delta}(z')\epsilon^{\mu\nu\alpha\beta}
\Delta_{\nu\gamma\delta}[z(\tau_1,s)-z(\tau',s')]=\nonumber\\
&&\quad\quad={1\over2}\sqrt{g(z)}t_{\alpha\beta}(z)\int\limits_{S(C)}dS_{\gamma\delta}(z')
\epsilon^{\mu\nu\alpha\beta}\Delta_{\nu\gamma\delta}(z-z').
\end{eqnarray}
We have consequently determined that
\begin{equation}
\frac{\delta A}{\delta x_\mu(\tau)}=\dot{x}_\alpha(\tau)
\int\limits_{S(C)}dS_{\gamma\delta}(z')\Delta_{\alpha\mu,\gamma\delta}^{(2)}
[x(\tau)-z(\tau',s')]
\end{equation}

In order to find the area derivative, {\it cf.} Eq.(3), we need to
calculate the second functional derivative of $A$,  at the points
$x(\tau_1)=x\left(\tau+{h\over2}\right)$ and
$x(\tau_2)=\left(\tau-{h\over2}\right)$. From the definition of
the area derivative we also surmise that only terms $\sim
\delta'(h)$ are relevant. Accordingly, it is straightforward to
surmise that
\begin{equation}
\frac{\delta A}{\delta
\sigma_{\mu\nu}(x(\tau))}=\int\limits_{S(C)}dS_{\gamma\delta}(z')\Delta^{(2)}_{\mu\nu,\gamma\delta}[x(\tau)
-z(\tau',s')]
\end{equation}
It is also easy to determine that
\begin{equation}
\frac{\delta}{\delta\sigma_{\mu_2{\nu_2}}(x(\tau_2))}\,\frac{\delta}{\delta\sigma_{\mu_1{\nu_1}}(x(\tau_1))}A=
\Delta^{(2)}_{{\mu_2{\nu_2}},{\mu_1{\nu_1}}}[x(\tau_2)-x(\tau_1)],
\end{equation}
while all higher derivatives give null contribution.

On the basis of the above analysis we determine
\begin{eqnarray}
&&\Phi^{(j)}[C]=1-\frac{iL}{2}
 \int\limits_0^1d\tau_1\int\limits_{S(C)}
dS\cdot\Delta^{(2)}(z-x_1)\cdot J+\nonumber\\
&+&\left(\frac{iL}{2}\right)^2
\int\limits_0^1{d\tau_2}\int\limits_0^{\tau_2}d\tau_1\left[
-J\cdot \Delta^{(2)}(x_2-x_1)\cdot J+\int\limits_{S(C)}
 dS\cdot\Delta^{(2)}(z-x_2)\cdot J
 \int\limits_{S(C)}
 dS\cdot\Delta^{(2)}(z-x_1)\cdot J\right]+\nonumber\\
&+&\left(\frac{iL}{2}\right)^3
\int\limits_0^1{d\tau_3}\int\limits_0^{\tau_3}d\tau_2\int\limits_0^{\tau_2}d\tau_1
\left[J\cdot\Delta^{(2)}(x_2-x_1)\cdot J\right.
\int\limits_{S(C)}dS\cdot\Delta^{(2)}(z-x_3)\cdot J+\nonumber\\
&+& J\cdot\Delta^{(2)}(x_3-x_2)\cdot
J\int\limits_{S(C)}dS\cdot\Delta^{(2)}(z-x_1)\cdot
J-\int\limits_{S(C)}dS\cdot\Delta^{(2)}(z-x_3)\cdot
J\times\nonumber\\
&\times&\int\limits_{S(C)}dS\cdot\Delta^{(2)}(z-x_2)\cdot
J\int\limits_{S(C)}dS\cdot\Delta^{(2)}(z-x_1)\cdot
J+\cdot\cdot\cdot.\left.\right]
\end{eqnarray}
In the above expression we have omitted terms in which the field
strength correlator $\Delta^{(2)}$ depends on distances between
two non-successive points. For example, the last line of the above
equation does not include the correlator
$\Delta^{(2)}(x(\tau_3)-x(\tau_1))$. The reason is, that such a
correlator [14,15,17,25] is assumed to behave as
$f\left(\frac{\mid x_3-x_1\mid^2}{T^2_g}\right)\simeq
f\left(\dot{x}^2\frac{(\tau_3-\tau_1)^2}{T_g^2}\right)$, which, in
turn, means that its contribution is suppressed by powers of
$T^2_g\sqrt{\Delta}$. Accordingly we obtain the following
expression for the spin factor:
\begin{equation}
\Phi^{(j)}[C]=P\exp\left[-{iL\over2}\int\limits_0^1d\tau
\int\limits_{S(C)}dS\cdot\Delta^{(2)}(z-x)\cdot
J+{L^2\over4}\int\limits_0^1d\tau_2\int\limits_0^{\tau_2}d\tau_1
J\cdot\Delta^{(2)}(x_2-x_1)\cdot J\right].
\end{equation}

We have arrived at a result of considerable interest for our
purposes, which we would like to analyze further. Let us start
from the second term in the exponent which is of special interest
[26] and represents the interaction of the quark color-magnetic
moment with the non-abelian background. In particular, let us
refer to the representation (28) of the two-point correlator,
which we rewrite in the form
\begin{equation}
\Delta^{(2)}_{\mu\nu,\lambda\rho}(\bar{x})=(\delta_{\mu\lambda}\delta_{\nu\rho}-
\delta_{\mu\rho}\delta_{\nu\lambda})(D+D_1)+(\bar{x}_\mu\bar{x}_\lambda\delta_{\nu\rho}-\bar{x}_\mu\bar{x}_\rho
\delta_{\nu\lambda}-(\mu\leftrightarrow\nu))D'_1.
\end{equation}
where we have denoted
$D'_1=\frac{\partial}{\partial\bar{x}^2}D_1(\bar{x})$.

Using expression (45) we find
\begin{equation}
J_{\mu\nu}\Delta^{(2)}_{\mu\nu,\lambda\rho}(\bar{x})J_{\lambda\rho}=2J^2(D+D_1)+4(J\cdot\bar{x})^2D'_1.
\end{equation}
For the case in hand we have that
$J_{\mu\nu}={i\over4}[\gamma_\mu,\gamma_\nu]$, so we can easily
determine that
\begin{equation}
J_{\mu\nu}\Delta^{(2)}_{\mu\nu,\lambda\rho}(\bar{x})J_{\lambda\rho}=6(D+D_1)+3\bar{x}^2D'_1.
\end{equation}
Expression (47) is positive definite and is associated [27] with
the ghost tachyonic pole which appears in the fermionic, or the
gluonic, propagator -an issue we shall not discuss further in this
paper.

With the help, now, of result (41) the first term in the
exponential (44) can be recast into the form
\begin{equation}
\int\limits
_0^1d\tau\int\limits_{S(C)}dS\cdot\Delta^{(2)}(z-x)\cdot
J=\int\limits _0^1d\tau J_{\mu\nu}\frac{\delta
A[C]}{\delta\sigma_{\mu\nu}(x(\tau))}.
\end{equation}
It is convenient, for the applications we have in mind, to rewrite
the above relation, by referring to the variation of ${ A}[C]$ as
it has been computed in Eq.(40):
\begin{equation}
\frac{\delta  A[C]}{\delta x_\mu(\tau)}\equiv g_\mu[x(\tau)].
\end{equation}
The above function is reparametrization invariant, thus
$\dot{x}_\mu g_\mu=0$.

Taking into account the result displayed in Eq.(40) we can write
\begin{equation}
\dot{x}_\mu\frac{\delta
A[C]}{\delta\sigma_{\mu\nu}(x(\tau))}=g_\nu[x(\tau)].
\end{equation}
An obvious solution of the above equation is
\begin{equation}
\frac{\delta A[C]}{\delta\sigma_{\mu\nu}(x)}={1\over\dot{x}^2}
[\dot{x}_\mu g_\nu(x)-\dot{x}_\nu g_\mu(x)].
\end{equation}
It can also be shown [28] that the above is the only possible
solution. The proof follows, basically, dimensional arguments:
Making the change $x\to \lambda x$ and $\tau\to\lambda\tau$, the
area derivative of $ A[C]$ scales like ${1\over\lambda^2}$. The
same scaling behavior goes for the function $g$, as can be easily
concluded from Eq.(40). Thus, the term appearing in (51) has the
right scaling properties. Any other term must be an antisymmetric
combination $K_{\mu\nu}$ which scales as ${1\over\lambda^2}$ and
is perpendicular to the velocity, {\it i.e.}, $\dot{x}_\mu
K_{\mu\nu}=0$. However,
 such a combination
expressible in terms of the boundary $x(\tau)$ cannot be found. It
thereby follows that the first term in the exponential which
defines the spin factor reads
\begin{equation}
\int\limits_0^1d\tau\int\limits_{S(C)}dS\cdot
\Delta^{(2)}(z-x)\cdot J=\int\limits_0^1d\tau\frac{\dot{x}_\mu
g_\nu-\dot{x}_\nu g_\mu}{\dot{x}^2}J_{\mu\nu}.
\end{equation}

\vspace*{.5cm}

{\bf 3. Loop Equations, Bianchi Identity.}

\vspace*{.5cm}

In this section we shall proceed to assess the capacity of the SVM
to expedite non-perturbative investigations in QCD by examining
whether Eq.(17), as formulated within the framework of the
stochastic approach,  satisfies the Polyakov/ Makeenko-Migdal
equations formulated in loop space. The latter, constitute the
most credible proposal for achieving a non-perturbative casting of
the theory, equivalently, one which provides a solid basis for
conducting non-perturbative investigations within its
framework\footnote{One might consider this approach as the
continuum space casting of lattice gauge theories, which defines
the `going' standard for such investigations.}. In addition, we
shall explicitly demonstrate the validity of the Bianchi identity.
We claim that these properties {\it must} be satisfied, if one is
to assert that the expression for the Wilson loop, as given in Eq.
(17), is to constitute a credible approximation to the full
theory.

Now, the loop equation for a contour without self-intersections
can be stated [6] as follows
\begin{equation}
\partial_\mu^{x(\tau)}\frac{\delta}{\delta\sigma_{\mu\nu}(x(\tau))}W[C]=\lim\limits_{\epsilon\to 0}
\int\limits_{\tau-\epsilon}^{\tau+\epsilon}d\tilde{\tau}
\frac{\delta}{\delta x_{\mu}(\tilde{\tau})}
\frac{\delta}{\delta\sigma_{\mu\nu}(x(\tau))}W[C]=0.
\end{equation}
Its verification constitutes the first, as well as simplest, test
the SVM must pass. To this end, let us insert Eq.(28) into Eq.(41)
whereupon, using the fact that the boundary is a closed contour,
we determine
\begin{eqnarray}
\frac{\delta}{\delta\sigma_{\mu\nu}(x(\tau))}A&=&\int
d^2\xi'\epsilon^{ab}\partial_a z_\mu(\xi')
\partial_bz_\nu(\xi')D[z(\xi')-x(\tau)]+\nonumber\\
&+&{1\over2}\int_0^1d\tau'[\dot{x}_\mu(\tau')(x_\nu(\tau')-x_\nu(\tau))-(\mu\leftrightarrow\nu)]D_1[x(\tau')-x(\tau)].
\end{eqnarray}

Our next step is to take the functional derivative of the above
equation. Our task becomes relatively easy as we notice, from
Eq.(53), that we only need those terms which contain the delta
function $\delta(\tilde{\tau}-\tau)$. Accordingly, we obtain
\begin{eqnarray}
&&\frac{\delta}{\delta x_\mu(\tilde{\tau})}\frac{\delta
A[C]}{\delta\sigma_{\mu\nu}(x(\tau))}=\int
d^2\xi'\epsilon^{ab}\partial_a
z_\mu(\xi')\partial_bz_\nu(\xi')\frac{\delta}{\delta
x_\mu(\tilde{\tau})}   D(z(\xi')-x(\tau))+\nonumber\\
&+&{1\over2}\int\limits_0^1d\tau'[\dot{x}_\mu(\tau')(x_\nu(\tau')-x_\nu(\tau))-(\mu\leftrightarrow\nu)]
\frac{\delta}{\delta x_\mu(\tilde{\tau})}D_1(x(\tau')-x(\tau))+\nonumber\\
&+&{3\over2}\delta(\tau-\tilde{\tau})\int\limits_0^1d\tau'\dot{x}_\nu(\tau')D_1(x'-x).
\end{eqnarray}

We now write
\begin{equation}
\frac{\delta}{\delta x_\mu(\tilde{\tau})}D(z(\xi')-x(\tau))=
\delta(\tau-\tilde{\tau}) \frac{\partial D(z-x)}{\partial
x_\mu}=-\delta(\tau-\tilde{\tau})\frac{\partial D(z-x)}{\partial
z_\mu}
\end{equation}
and  similarly for the derivative of $D_1$. Thus
\begin{eqnarray}
&&\partial^{x(\tau)}_\mu\frac{\delta
A[C]}{\delta\sigma_{\mu\nu}(x(\tau))}=-\int
d^2\xi'\epsilon^{ab}\partial_a z_\mu(\xi')\partial_b
z_\nu(\xi')\frac{\partial}{\partial z_\mu}D(z-x)-\nonumber\\
&-&{1\over2}\int\limits_0^1d\tau'
[\dot{x}_\mu(\tau')(x_\nu(\tau')-x_\nu(\tau))-(\mu\leftrightarrow\nu)]
\frac{\partial}{\partial x'_\mu}D_1(x'-x)+{3\over2}\oint
d\bar{x}_\nu D_1(\bar{x}).
\end{eqnarray}

 Since the boundary is a (closed) loop, we conclude that the first term takes
the form
\begin{eqnarray}
\int d^2\xi'\epsilon^{ab}\partial_a
z_\mu(\xi')\partial_bz_\nu(\xi')
\frac{\partial}{\partial z_\mu}D(z-x)&=&\int d^2\xi'\epsilon^{ab}\partial_aD(z-x)\partial_bz_\nu(\xi')=\nonumber\\
&=&-\int\limits_0^1d\tau'\dot{x}_\nu(\tau')D(x'-x)=-\oint_Cd\bar{x}_\nu
D(\bar{x}).
\end{eqnarray}
With the same reasoning we have for the second term
\begin{eqnarray}
&& \int\limits_0^1d\tau' \left[\dot{x}_\mu(\tau')
\left(x_\nu(\tau')-x_\nu(\tau)\right)-(\mu\leftrightarrow\nu)\right]
\frac{\partial}{\partial x'_\mu}D_1(x'-x)=\nonumber\\
&=&-\int\limits_0^1d\tau'\dot{x}_\nu(\tau')(x_\mu(\tau')-(x_\mu(\tau))
\frac{\partial}{\partial
x'_\mu}D_1(x'-x)=-2\oint_Cd\bar{x}_\nu\mid\bar{x}\mid^2\frac{\partial}{\partial
\mid\bar{x}\mid^2}D_1(\bar{x}).
\end{eqnarray}

 Accordingly,
\begin{equation}
\partial^{x(\tau)}_\mu\frac{\delta
A[C]}{\delta\sigma_{\mu\nu}(x(\tau))}=-\oint_Cd\bar{x}_\nu
\left[D(\bar{x})+{3\over2}D_1(\bar{x})+\mid\bar{x}\mid^2\frac{\partial}{\partial\mid\bar{x}\mid^2}
D_1(\bar{x})\right].
\end{equation}
Given that the functions $D$ and $D_1$ depend only on the distance
$\mid\bar{x}\mid$, the right hand side of the above equation
vanishes and the loop equation is satisfied. In fact what {\it is
known} is that the minimal area satisfies the Makeenko-Migdal
equation aymptotically. The above result, on the other hand, can
be considered as new in the sense that it is an {\it exact result}
in the framework of the Stochastic Vacuum hypothesis.

Our next concern is to confirm the Bianchi Identity (BI) in the
framework of the SVM. The reason we are interested in such a
confirmation is based on the fact that the zigzag, or
backtracking, symmetry characterizes the Wilson loop functional:
It is invariant under reparametrizations of the form $x(\tau)\to
x(\alpha(\tau))$, even if $\alpha'<0$. It can be shown [24] that a
Stokes type functional, of which the Wilson loop is a prime
example, which respects the aforementioned symmetry also satisfies
Eq. (10), which can be written in the form
\begin{equation}
\epsilon^{\kappa\lambda\mu\nu}\partial^{x(\tau)}_\lambda\frac{\delta}{\delta\sigma_{\mu\nu}(x(\tau))}W[C]
={1\over
N_c}\epsilon^{\kappa\lambda\mu\nu}\left\langle\bigtriangledown_\lambda
F_{\mu\nu}\exp\left(i\oint_Cdx\cdot {\cal
A}\right)\right\rangle_{\cal A}=0.
\end{equation}
The above relation forms the bridge between the zigzag symmetry
and the BI in the framework of a gauge field theory.

As it is, now obvious the  area
 $S=\int d^2w\sqrt{g}$  does not respect the zigzag symmetry and in this
sense $e^{-\sigma S}$ is not a good representative for $W[C]$. On
the other hand, the action $A[C]$ in the stochastic approximation
{\it is} invariant under zigzag parametrizations, but due to the
truncation $e^{-A}$ it is not obvious that it is a Stokes type
functional. We are thereby obliged to confirm that the BI is
explicitly satisfied and that, consequently, the stochastic
approximation produces Stokes type functionals. To this end we
first observe, using Eq. (41), that
\begin{equation}
\epsilon^{\kappa\lambda\mu\nu}\partial_\lambda^{x(\tau)}\frac{\delta}{\delta\sigma_{\mu\nu}(x(\tau))}
\ e^{-A[C]}
=\left(\epsilon^{\kappa\lambda\mu\nu}\int\limits_{S(C)}dS_{\alpha\beta}(z)\partial_\lambda
\Delta^{(2)}_{\mu\nu,\alpha\beta}(z-x)\right)e^{-A[C]}.
\end{equation}

Using Eq.(29) we find that
\begin{equation}
\epsilon^{\kappa\lambda\mu\nu}\int\limits_{S(C)}dS_{\alpha\beta}(z)\partial_\lambda
\Delta^{(2)}_{\mu\nu,\alpha\beta}(z-x)=2X_\kappa,
\end{equation}
where we have set
\begin{equation}
X_\kappa=\int\limits_{S(C)}dS_{\alpha\beta}\Delta_{\kappa\alpha\beta}(z-x).
\end{equation}
Now, we have established that the surface over which we integrate
the correlators is determined by Eq. (27). Multiplying that
equation with  $\epsilon^{\sigma\kappa'\mu'\nu'}$ we determine
\begin{equation}
t_{\mu\nu}X_\kappa+t_{\nu\kappa}X_\mu+t_{\kappa\mu}X_\nu=0.
\end{equation}
The above equations form a homogeneous system whose only solution
is $X_\kappa=0$, a result which, together with Eqs. (61) and (62)
confirms the validity of the BI within the framework of the SVM.

\vspace*{.5cm}

{\bf 4. Meson-Meson Scattering.}

\vspace*{.5cm}

 In addition to confinement, which constitutes a profoundly non-perturbative
 problem, there do exist specific {\it dynamical} processes, whose
theoretical confrontation also calls for non-perturbative methods
of analysis.
 One such situation arises in connection with the theoretical description of
high energy scattering amplitudes for which the soft sector of the
theory is involved. From the experimental point of view, one such
case arises in connection with Regge kinematics, entering directly
the theoretical description of, among others, diffractive and
low-x physics processes. In this section we shall study a
simulated case of a meson-meson scattering process whose
quark-based description is of the general form:
\[
(1\bar{1})+(2\bar{2})\to(3\bar{3})+(4\bar{4})
\]
Adopting a standard picture, already employed in the QCD
literature (see, for example [29-31]) according to which quark 1
from the first meson and antiquark $\bar{2}$ from the second meson
are very heavy, in comparison to the incoming total energy, hence
their worldlines are considered to remain intact from the gluon
field action and can be described in the framework of the eikonal
approximation. The light pairs $\bar{1},2$ and $\bar{3},4$, on the
other hand, are annihilated and produced in the $t$-channel, where
the eikonal approximation is not valid and a full treatment is
called for their description. In the Worldline framework the
process is schematically pictured in space-time by the straight
eikonal lines $(1\to 3)$ and $(\bar{2}\to\bar{4})$, describing an
intact quark and anti-quark and by the curves $(\bar{1}\to 2)$ and
$(\bar{3}\to 4)$ which correspond, respectively, to the
annihilated and produced quark antiquark pairs. The structure of
the  field theoretical amplitude can be written as follows (see
Fig.):
\begin{equation}
G(x_4,x_3,x_2,x_1)=\langle iS_F(x_4,x_3\mid {\cal A})
iS_F(x_3,x_1\mid {\cal A})iS_F(x_1,x_2\mid {\cal
A})iS_F(x_2,x_4\mid {\cal A})\rangle_{\cal A}.
\end{equation}
In the above expression $iS_F$ is the full fermionic propagator
which, in the framework of the Worldline formalism, assumes the
form [16]
\begin{equation}
iS_F(y,x\mid {\cal A})=\int_0^\infty dL\, e^{-Lm^2}
\int\limits_{{\stackrel
{x(0)=x}{x(L)=y}}}Dx(\tau)e^{-{1\over4}\int\limits_0^Ld\tau\dot{x}^2}
\left[m-\frac{\gamma\cdot\dot{x}(L)}{2}\right]\Phi^{(1/2)}(L,0)P\exp\left(i\int\limits_0^Ld\tau\dot{x}\cdot
\cal{A}\right),
\end{equation}
where $\Phi^{(j)}$ is the so-called spin factor for the matter
particles entering the system. For us, it means that
$j={1\over2}$.

Inserting the above formula into Eq.(66) we find
\begin{eqnarray}
&&G(x_4,x_3,x_2,x_1)=\prod\limits_{i=1}^4\int\limits_0^\infty
d\tau_i\theta(\tau_i-\tau_{i-1})e^{-(\tau_i-\tau_{i-1})m_i^2}
\int\limits_{{\stackrel {x(0)=x_4}{x(\tau_4)=x_4}}}\int
Dx(\tau)\delta[x(\tau_3)-x_3]\delta[x(\tau_2)-x_1]\times\nonumber\\
 &&\times\delta[x(\tau_1)-x_2]
\exp\left[-{1\over4}\int\limits_0^{\tau_4}
d\tau\,\dot{x}^2(\tau)\right]({\rm spin
\,\,structure})\left\langle P\exp\left(i\oint_C dx\cdot {\cal
A}\right)\right\rangle_{\cal A},
\end{eqnarray}
where the term $spin\,\, structure$ corresponds to the following
expression
\begin{equation}
({\rm spin\,\, structure})
=\prod\limits_{i=4}^1\left[m_i-{1\over2}\gamma\cdot{\dot{x}}(\tau_i)\right]\Phi^{(1/2)}(\tau_i,\tau_{i-1}),\,\,(\tau_0\equiv
0).
\end{equation}

In principle, the Wilson loop appearing in Eq.(68) incorporates
the dynamics (perturbative, as well as non-perturbative) of the
process. In the framework of the SVM it assumes the form

\begin{equation}
\left\langle P\exp\left(i\oint_C dx\cdot {\cal
A}\right)\right\rangle_{\cal A}=\exp\left[-{1\over
2}\int\limits_{S(C)}dS_{\mu\nu}(z)\int\limits_{S(C)}dS_{\lambda\rho}(z')\Delta^{(2)}_{\mu\nu,
\lambda\rho}(z-z')\right]\equiv e^{-A[C]}.
\end{equation}

In the present Section we are going to calculate the amplitude
(68), using the above expression which gives the structure of the
Wilson loop in the framework of the SVM. The particular method to
be adopted is a kind of a $``$semiclassical" approximation based
on a combined minimization of the action $ A[C]$- see Eq.(39)-
with respect to the surface $S[C]$ and of the surface $S[C]$ with
respect to the boundary $C$. The reasoning behind this procedure
is that, according to Eq.(68), in order to obtain the full
amplitude it does not suffice to determine the minimal surface
bounded by a given specific contour but one needs to proceed even
further and sum over all possible boundaries with a weight of the
form
\begin{equation}
S[x]={1\over4}\int\limits_0^{\tau_4}d\tau\,\dot{x}^2+ A[C].
\end{equation}

The above described approximation will allow us to determine the
dominant contribution to the Worldline integral (68) in the
stochastic limit $T_g^2\sqrt{\Delta}\ll 1$.

\vspace*{.2cm}

The
variation, $g_{\mu}[x(\tau)]$, of $ A[C]$ under changes of the
boundary is given by Eq. (39). Accordingly, the correlator
contributions become stationary for the ``classical" trajectory
\begin{equation}
g_\mu[x_{{\rm cl}}]=0.
\end{equation}
Using the expansion for the correlator according to Eq.(28) it is
easy to see that
\begin{equation}
g_\mu[x(\tau)]=2\dot{x}_\alpha(\tau)
R_{\alpha\mu}[x(\tau)]-{1\over2}\dot{x}_\alpha(\tau)Q_{\alpha\mu}[x(\tau)],
\end{equation}
with
\begin{equation}
R_{\alpha\mu}[x(\tau)]=\int\limits_{S(C)}dS_{\alpha\mu}(z')D[x(\tau)-z(s',\tau')]
\end{equation}
and
\begin{equation}
Q_{\alpha\mu}[x(\tau)]=\int
d\tau'\left[\dot{x}_\mu(\tau')\left(x_\alpha(\tau')-x_\alpha(\tau)\right)-(\mu\leftrightarrow
\alpha)\right]D_1[x(\tau)-x(\tau')].
\end{equation}

It is worth noting that the above expressions are
reparametrization invariant. Also, in the last relation the
integration covers the whole range of the $\tau$ variable.
Following Refs. [30,31] the minimal surface bounded by two
infinite rods at a relative angle $\theta$, has (in
four-dimensional Euclidean space) the shape of a
(three-dimensional) helicoid, which is the only surface that can
be spanned by straight lines [32]. In the considered process the
eikonal lines $1\to3,\,\bar{2}\to\bar{4}$, play the role of the
$``$rods", while the angle $\theta$ is connected, via analytic
continuation [33], to the logarithm of their total energy $s$.

Given the above specifications, consider the following, helpful,
parametrization of the boundary $C$: For $0<\tau<\tau_1$ we have a
straight line segment, $x^{(1)}$, going from the point $x_4$ to
the point $x_2$. Introducing, moreover, for convenience the {\it
length} $2T=\mid x_4-x_2\mid$ and reparametrizing according to
$\tau\to\frac{2T}{\tau_1}\tau-T$, we write
\begin{equation}
x_\mu^{(1)}=(\tau,0,0,0), \,\,-T<\tau<T,
\end{equation}
with  $x_\mu^{(1)}(-T)=x_4,\, x_\mu^{(1)}(T)=x_2$.

The second eikonal line $x^{(3)}(\tau),\,\tau_2<\tau<\tau_3$, goes
from the point $x_1$ to the point $x_3$ at a relative angle
$\theta$ with respect to $x^{(1)}$, while a distance $b$ (impact
parameter) separates the two linear contours in a transverse
direction. Introducing the distance $2T_1=\mid x_3-x_1\mid$ and
reparametrizing according to
\begin{equation}
\tau\to
T_1\left(\frac{2}{\tau_3-\tau_2}\tau-\frac{\tau_3+\tau_2}{\tau_3-\tau_2}\right)
\end{equation}
we write
\[
x_\mu^{(3)}(\tau)=(-\tau\cos
\theta,\,-\tau\sin\theta,\,b,\,0),\quad -T_1<\tau<T_1,
\]
with $x_\mu^{(3)}(-T_1)=x_1,\,x_\mu^{(3)}(T_1)=x_3.$

In the following we shall assume, just for convenience, that
\[
2T=\mid x_4-x_2\,\mid\sim \,\mid x_3-x_1\mid=2T_1.
\]
For $\tau_1<\tau<\tau_2$, we have a helical curve
$x_\mu^{(2)}(\tau)$, which joins the points
$x_2=x_\mu^{(2)}(\tau_1)$ and $x_1=x_\mu^{(2)}(\tau_2)$,
representing the exchanged light quarks. Performing, now, the
change
\newline
$\sigma=\frac{b}{\tau_2-\tau_1}(\tau-\tau_1)$, we write
\begin{equation}
x_\mu^{(2)}(\sigma)=\left(\phi(\sigma)\cos\frac{\theta
\sigma}{b},\, \phi(\sigma)\sin\frac{\theta
\sigma}{b},\sigma,\,0\right),\,\,0<\sigma<b.
\end{equation}

The continuity of the boundary requires
\[
x_\mu^{(1)}(T)=x_\mu^{(2)}(0)=x_2\quad {\rm
and}\,\,x_\mu^{(2)}(b)=x_\mu^{(3)}(-T)=x_1,
\]
or
\begin{equation}
\phi(0)=\phi(b)=T.
\end{equation}
The final helical curve is $x^{(4)}(\tau)$, which, for
$\tau_3<\tau<\tau_4$, joins the points $x_3=x^{(4)}(\tau_3)$ and
$x_4=x^{(4)}(\tau_4)$. Making one more, final, reparametrization,
namely $\sigma=\frac{b}{\tau_4-\tau_3}(\tau-\tau_3)$ we write
\begin{equation}
x_\mu^{(4)}(\sigma)=\left(-\phi(\sigma)\cos {\theta \sigma\over
b},\,-\phi(\sigma)\sin {\theta \sigma\over b},\,\sigma,
0\right),\,\,0<\sigma<b.
\end{equation}
Once again, Eq.(79) takes care of the continuity of the boundary.
Now, the minimal surface is bounded by the (four) curves specified
by Eqs.(76)-(80) and can be spanned by straight lines parametrized
as follows
\begin{equation}
z_\mu(\xi)=\frac{T-\tau}{2T}x_\mu^{(4)}(\sigma)+\frac{T+\tau}{2T}x_\mu^{(2)}(\sigma)
=\left({\tau\over T}\phi(\sigma)\cos{\theta \sigma\over
b},\,{\tau\over T}\phi(\sigma)\sin{\theta \sigma\over
b},\,\sigma,0\right).
\end{equation}
It can be easily proved that the surface defined by the above
equation is minimal, irrespectively of the function $\phi$:
\begin{equation}
\partial_\tau\left[\frac{(\dot{z}\cdot
z')z'_\mu-z'^2\dot{z}_\mu}{\sqrt{g}}\right]+
\partial_\sigma\left[\frac{(\dot{z}\cdot
z')\dot{z}_\mu-\dot{z}^2z'_\mu}{\sqrt{g}}\right]=0.
\end{equation}

One observes that the minimization of the surface is not enough
for the complete specification of the parametrization of the
helicoid. Accordingly, we go back to Eq.(72), which determines the
boundary that dominates the path integration (68). A first
observation is that, due to the antisymmetric nature of
$R_{\alpha\mu}$ and $Q_{\alpha\mu}$, the function $g_\mu$ vanishes
when $x_\mu(\tau)$ represents a straight line. Thus Eq.(72) is
trivially satisfied for the eikonal sector of the boundary.
Non-trivial contributions are coming only from the helices
$x_\mu^{(2)}$ and $x_\mu^{(4)}$. One can simplify Eq.(73) by
computing the leading behavior of the functions $R_{\alpha\mu}$
and $Q_{\alpha\mu}$ using the fact that the functions $D$ and
$D_1$, as defined in the SVM scheme -and measured in lattice
calculations [25]- decay exponentially fast for distances which
are large in comparison with the correlation length $T_g$. In this
connection and upon writing
 \[
 x(\sigma')=x(\sigma)+(\sigma'-\sigma)\dot{x}(\sigma)+{1\over2}(\sigma'-\sigma)^2\ddot{x}(\sigma)+\cdot\cdot\cdot,
 \]
 we find, for the second term in Eq.(73),
\begin{eqnarray}
\dot{x}_\alpha Q_{\alpha\mu}&=& {1\over
2}\left[(\dot{x}^2)\ddot{x}_\mu-(\dot{x}\cdot\ddot{x})\dot{x}_\mu\right]
\int\limits_0^b
d\sigma'(\sigma'-\sigma)^2D_1\left[\dot{x}^2\frac{(\sigma'-\sigma)^2}{T^2_g}\right]
+\cdot\cdot\cdot=\nonumber\\
&=&{1\over \mid
\dot{x}\mid}\left(\ddot{x}-\frac{\dot{x}\cdot\ddot{x}}{\dot{x}^2}\right)\frac{1}{T_g\alpha_1}+\cdot\cdot\cdot,
\end{eqnarray}
where\footnote{We have omitted terms suppressed by powers of
$T_g^2$}
\[ \frac{1}{\alpha_1}\equiv T^4_g\int\limits_0^\infty
dw\,w^2D_1(w^2).
\]

In the last expression we have used, as in Eqs.(33) and (34), the
dimensionless parameter $w= \mid z \mid /T_g$.
 It must be noted
that the coefficient $1/{\alpha_1}$ is a small number. Taking into
account that $T_g$ is of the order of 0.1fm and the string tension
(see Eq.(33)) $\sigma\simeq0.18$ GeV$^2$ it is readily seen that
$1/{\alpha_1}=O(\sigma T_g^2)$.

 Noting that
\begin{eqnarray}
&& z_\mu(\sigma,\tau=T)= x_\mu^{(2)}(\sigma),\quad
z_\mu(\sigma,\tau=-T)=x_\mu^{(4)}(\sigma)\nonumber\\
&&\partial_\tau
z_\mu(\sigma,\tau)=\dot{z}_\mu(\sigma,\tau)={1\over
2T}[x_\mu^{(2)}(\sigma)-x_\mu^{(4)}(\sigma)],
\end{eqnarray}
the leading behavior of the first term of the rhs of (73) can be
easily determined. One finds
\begin{eqnarray}
\dot{x}_\alpha
R_{\alpha\mu}&=&{1\over2}\dot{x}^2\left(\dot{z}_\mu-
\frac{(\dot{x}\cdot\dot{z})}{\dot{x}^2}{\dot{x}_\mu}\right)\int\limits_{-T}^T
d\tau'\int\limits_0^b d\sigma'\,D\left[\dot{x}^2
\frac{(\sigma'-\sigma)^2}{T_g^2}\right]+\cdot\cdot\cdot\nonumber\\
&=&2T\mid\dot{x}\mid\left(\dot{z}_\mu-\frac{(\dot{x}\cdot\dot{z})}{\dot{x}^2}\dot{x}_\mu\right){\mu^2\over
T_g}+\cdot\cdot\cdot,
\end{eqnarray}
where we have introduced the parameter
\begin{equation}
\mu^2\equiv T^2_g\int\limits_0^\infty dw\,D(w^2)\sim O(\sigma).
\end{equation}
Thus, the function $g$ takes, to leading order, the form
\begin{equation}
g_\mu=\frac{1}{\mid\dot{x}\mid T_g}\left[4T\mu^2\dot{x}^2
\left(\dot{z}_\mu-\frac{\dot{x}\cdot\dot{z}}{\dot{x}^2}\dot{x}_\mu\right)
-\frac{1}{2\alpha_1}\left(\ddot{x}_\mu-\frac{\dot{x}\cdot\ddot{x}}{\dot{x}^2}\dot{x}_\mu\right)\right].
\end{equation}

Now, we recall from its definition that the $g$-function provides
a measure of the change of $A[C]$ when the Wilson contour is
altered as a result of some interaction which reshapes its
geometrical profile. In this sense, it contains important
information concerning the dynamics of the problem under study.
The structure of the $g$-function, as it appears in the above
equation, is quite general and exhibits its dependence, not only
on the boundary but on the minimal surface as well. It is worth
noting that this fact is strictly associated with the non-Abelian
nature of the theory since the function $D$ -and consequently
$\mu^2$- disappears [14,15] in an Abelian gauge theory.

Taking into account that for the helicoids parametrization the
velocity $\dot{x}$ has three non-zero components, while $\ddot{x}$
and $\dot{z}$ have only two, we conclude that Eq. (72) can be
satisfied only if
\begin{equation}
4T\mu^2\dot{x}^2\dot{z}_\mu-{1\over 2\alpha_1}\ddot{x}_\mu=0.
\end{equation}
Inserting in Eq.(88) the helical parametrization one easily finds
that the function $\phi$ must be a constant. Taking, now, into
account Eq.(79) we determine this constant to be the length $T$.
It is then very easy to see that this result leads to the
conclusions
\begin{equation}
\dot{x}\cdot\dot{z}=0,\,\,\dot{x}\cdot\ddot{x}=0
\end{equation}
and
\begin{equation}
\dot{x}^2=-\frac{1}{8\mu^2\alpha_1}\frac{\theta^2}{b^2}=1+\frac{T^2\theta^2}{b^2},
\end{equation}

This equation cannot be satisfied in Euclidean space. In Minkowski
space the angle $\theta$ becomes imaginary
$\theta\to-i\chi\simeq-i\ln\left({s\over m^2}\right)$ ($s$ is the
total energy of the $``$heavy" quarks that form the $``$rods" and
$m(\simeq m_1\simeq m_3)$ their mass and Eq.(90) has a positive
definite solution:
\begin{equation}
\frac{T^2\chi^2}{b^2}=1-\frac{1}{8\mu^2\alpha_1}\frac{\chi^2}{b^2}.
\end{equation}
In the last equation we have analytically continued only the angle
between the $``$rods" and not the parameter $T$. Our reasoning is
based on the one presented in [31]: Eqs.(79) and (90) define a
contour of steepest descent for the path integration. For such a
contour the parameter $T$ is determined to be imaginary (the
formulation is still Euclidean and $\theta$ is real). After
performing the $T$ integration the only parameter remaining for
analytic continuation is the angle between the $``$rods". Eq.(91)
indicates that, equivalently, one can first analytically continue
the angle variable to imaginary values leaving the $T$ parameter
real. In any case it is obvious that the impact parameter must
grow with the incoming energy: $ T_g b\sim\ln s$, a conclusion
which is in agreement with the landmark result of Cheng and Wu
[34].

The preceding analysis obviously repeats itself for the two
helical curves
 $x^{(2)}$ and $x^{(4)}$ and has led us to a specific parametrization
 for the Wilson loop, which plays the dominant role in the path
 integration in Eq.(68). We are now in position to determine the leading contribution to
 the action (71):
\begin{equation}
S_{\rm cl}={1\over4} \int\limits ^{\tau_4}_0
d\tau\dot{x}^2_{cl}(\tau)+A[C]_{\rm cl}.
\end{equation}

Our first step is to expand the second term of the integrand in
powers of $T^2_g\sqrt{\Delta}$. The first term of such an
 expansion is the familiar Nambu-Goto string. The next term,
 which reveals the rich structure of the SVM, is the so-called
 $``$rigidity term", representing the extrinsic curvature of a
 surface embedded  in a four-dimensional [35] background:

\begin{equation}
A[C]=\sigma\int d^2\xi\sqrt{g}+{1\over\alpha_0}\int
d^2\xi\sqrt{g}g^{ab}\partial_a t_{\mu\nu}\partial_b
t_{\mu\nu}+\cdot\cdot\cdot,
\end{equation}

where $\sigma$ is the string tension as defined in Eq.(32).

The coefficient of the rigidity term reads
\begin{equation}
{1\over\alpha_0}\equiv{1\over32}T^4_g\int
d^2w\,w^2(2D_1(w^2)-D(w^2))\sim\ O(\sigma T_g^2).
\end{equation}

Terms proportional to $T_g^6$ entering the expansion in Eq.(93)
will be considered negligible in our analysis. We have also
omitted the term $\int d^2{\xi}\sqrt{g}R$, since in two dimensions
the curvature is a total derivative. Using the helicoids
parametrization (81), with $\phi=T$, the Nambu-Goto term in
Eq.(93) takes the form
\begin{equation}
\int d^2\xi\sqrt{g}=\int\limits_{-T}^Td\tau\int\limits_0^b
ds\sqrt{1+\frac{\tau^2\theta^2}{b^2}}
=bT\left[\sqrt{1+p^2}+{1\over
p}\ln\left(\sqrt{1+p^2}+p\right)\right],
\end{equation}
where $p=\frac{T\theta}{b}$.

To proceed further we analytically continue to Minkowski space
where we can use Eq.(91) to determine
\begin{equation}
bT\sqrt{1+p^2}\to bT\sqrt{1-\frac{T^2\chi^2}{b^2}}\simeq
b\left(1-\frac{1}{8\alpha_1\mu^2}\frac{\chi^2}{b^2}\right)^{1/2}\frac{1}{
\sqrt{8\alpha_1\mu^2}}\simeq \frac{b}{
\sqrt{8\alpha_1\mu^2}}+{\cal O}(T^3_g)
\end{equation}
and
\begin{equation}
\frac{bT}{p}\ln\left(\sqrt{1+p^2}+p\right)\to
\frac{bT}{-iT\chi/b}\ln\left[\sqrt{1-\frac{T^2\chi^2}{b^2}}
-i{T\chi\over b}\right]\simeq \frac{\pi
b^2}{2\chi}-\frac{b}{\sqrt{8\alpha_1\mu^2}}+ {\cal O}(T^3_g).
\end{equation}
Thus
\begin{equation}
\sigma\int d^2\xi\sqrt{g}\to\frac{\sigma\pi b^2}{2\chi}.
\end{equation}
In the same framework, the contribution of the rigidity term takes
the form
\begin{eqnarray}
\int d^2\xi\sqrt{g}\,g^{ab}\partial_a\,
t_{\mu\nu}\partial_b\,t_{\mu\nu} &=& \int\limits_{-T}^T
d\tau\int\limits_0^b ds
\frac{1}{\sqrt{1+\frac{\theta^2\tau^2}{b^2}}}\left(\frac{\theta^2}{b^2}+{1\over2}
\frac{\theta^4}{b^4}\tau^2\right)\nonumber\\&=&\theta
\left[{3\over2}\ln\left(\sqrt{1+p^2}+p\right)+{1\over2}p
\sqrt{1+p^2}\right].
\end{eqnarray}

It follows that in Minkowski space we have
\begin{equation}
{1\over\alpha_0}\int
d^2\xi\sqrt{g}g^{ab}\partial_at_{\mu\nu}\partial_ bt_{\mu\nu}\to
-\frac{3\pi}{4\alpha_0}\chi.
\end{equation}
For the full estimation of the classical action, {\it cf.}
Eq.(92), one should also take into account the presence of the
classical kinetic term. Non trivial contributions come from the
helical curves $x^{(2)}(\bar{1}\to 2)$ and $x^{(4)}(\bar{3}\to
4)$:
\begin{equation}
\frac{b}{4(\tau_2-\tau_1)}\int\limits_0^b ds(\dot{x}^{(2)})^2+
\frac{b}{4(\tau_4-\tau_3)}\int\limits_0^b ds(\dot{x}^{(4)})^2
=\frac{b^2\dot{x}^2}{4(\tau_2-\tau_1)}+\frac{b^2\dot{x}^2}{4(\tau_4-\tau_3)}.
\end{equation}

Now we have to take into account that both $\tau_2-\tau_1$ and
$\tau_4-\tau_3$ must be integrated with weights
$e^{-(\tau_2-\tau_1)m_0^2}$ and $e^{-(\tau_4-\tau_3)m_0^2}$,
respectively. These integrals, as it turns out, are dominated by
the values
$\tau_2-\tau_1=\tau_4-\tau_3=\frac{b\mid\dot{x}\mid}{2m_0}$,
leading to a final kinetic contribution of the form
\begin{equation}
2m_0b\mid\dot{x}\mid=2\frac{m_0}{\sqrt{8\alpha_1\mu^2}}\chi.
\end{equation}
Here, $m_0(\simeq m_2 \simeq m_4)$ is the (current) mass of the
light quarks, thus the result expressed by (102) can be considered
negligible.

From the above analysis we conclude that
\begin{equation}
S_{\rm cl}\approx\frac{\sigma\pi
b^2}{2\chi}-\frac{3\pi}{4\alpha_0}\chi
\end{equation}
Putting aside, for now, the possible corrections to $A[C]$ which
arise from fluctuations of the boundary as well as the spin factor
contribution, let us consider the result (103) as a whole, except
for terms $\sim \rm mass$. To obtain the final expression for the
scattering amplitude one must integrate over the impact parameter:
\begin{equation}
\int
d^2b\exp\left(i\vec{q}\cdot\vec{b}-\frac{\sigma\pi}{2\chi}b^2\right)\propto\exp\left(-{1\over2\pi\sigma}q^2\chi\right).
\end{equation}

Combining (103) and (104) we find, for the scattering amplitude, a
Regge behavior of the form $s^{\alpha'_R(0)t+\alpha_R(0)}$ with
\begin{equation}
\alpha'_R(0)={1\over2\pi\sigma}\,\,{\rm
and}\,\,\alpha_R(0)=\frac{3\pi}{4\alpha_0}.
\end{equation}

The form of the slope $a'_R(0)$, indicated in the last equation,
is the same as the one obtained in Ref.[32]. In that work the
authors have applied, in the framework of the SVM, a different
method based on the path integral-Hamiltonian duality and
consequently their result is confined in the region $t>0$. In this
sense our result extends the same value
$a'_R(0)=1/2\pi\sigma\simeq0.9$ GeV$^{-2}$  for all the values of
square momentum transfer.

It is well known [26,32] that the intercept $a_R(0)$ receives
significant contribution from non-perturbative corrections to
quark self-energy. We shall comment on this interesting issue in
the next Section. The result indicated in Eq.(105) does not take
into account the aforementioned corrections  thus it is very
sensitive to different lattice data or parametrizations. For
example, using Ref.[25], the coefficient $1/a_0$ of the rigidity
term is negative and one needs [32] the large (and negative) quark
self-energy corrections to restore the phenomenological value of
the intercept. On the other hand adopting a certain [17,36]
parametrization for the functions $D$ and $D_1$ (see Appendix B)
one obtains for the string tension the value $\sigma=0.175$
GeV$^2$ and for the coefficient of the rigidity term the value
$1/a_0=0.276$. With these numbers we obtain for the Reggeon slope
the value $a'_R(0)=0.91$GeV$^{-2}$ and for the Reggeon intercept
the value $a_R(0)=0.65$ in good agreement with the
phenomenological values $\alpha'_R(0)= 0.93$ GeV$^{-2}$ and
$\alpha_R(0)=0.55$  [37].

\vspace*{.5cm}

{\bf 5. Boundary Fluctuations and the Role of the Spin Factor}.
\vspace*{.5cm}

As repeatedly mentioned in our narration, corrections to the
amplitude (68), beyond semiclassical ones, are expected to arise
from fluctuations of the boundary of the surface on which the
two-point correlator $``$lives". Fluctuations of the surface
itself can be taken into account by higher order correlators.
This, in fact, is the big difference which distinguishes the SVM
approach, in comparison with Nambu-Goto type approaches.

We begin our related considerations by expanding the action (71)
around the helicoid classical solution:
\begin{eqnarray}
&&S=S_{{\rm
cl}}-{1\over2}\int\limits_0^{\tau_4}d\tau\,y(\tau)\ddot{x}^{{\rm
cl}}(\tau)+{1\over2}\int\limits_0^{\tau_4}d\tau\int\limits_0^{\tau_4}d\tilde{\tau}\,
y_\alpha\tau)\times\nonumber\\&&
\times\left[-{1\over2}\delta_{\alpha\beta}\frac{\partial^2}{\partial\tau^2}\delta(\tau-\tilde{\tau})
+\frac{\delta^2A[C]}{\delta x^{{{\rm cl}}}_\alpha(\tau)\delta
x_\beta(\tilde{\tau})}\right]y_\beta(\tilde{\tau})+\cdot\cdot\cdot,
\end{eqnarray}
where $y=x-x^{\rm cl}$.

 Using the results of Sections 2 and 3 one can easily determine that
\begin{eqnarray}
\frac{\delta^2A[C]}{\delta x_\alpha(\tau)\delta
x_\beta(\tilde{\tau})}=&&\dot{x}_\mu(\tau)\dot{x}_\nu(\tilde{\tau})
\Delta^{(2)}_{\mu\alpha,\nu\beta}[x(\tau)-x(\tilde{\tau})]-\nonumber\\&&
-\frac{\partial}{\partial\tau}\delta(\tau-\tilde{\tau})\int\limits_{S(C)}dS_{\lambda\rho}(z')
\Delta^{(2)}_{\alpha\beta,\lambda\rho}[z(\xi'-x(\tau)]+\nonumber\\&&
+\dot{x}_\alpha(\tau)\int
ds\,\alpha(\tilde{\tau},s)\dot{z}_\lambda(\tilde{\tau},s)z'_\rho(\tilde{\tau},s)\epsilon^{\kappa\nu\lambda\rho}
\Delta_{\kappa\alpha\mu}[z(\tilde{\tau},s)-x(\tau)],
\end{eqnarray}
where we have written
\[
\frac{\delta z_\mu(\tau,s)}{\delta
x_\nu}=\delta_{\mu\nu}\delta(\tau-\tilde{\tau})a(\tilde{\tau},s).
\]
The second term on the rhs of Eq.(107) is simply the area
derivative which, as we have seen in Section 3, has the general
form $\frac{\delta A[C]}{\delta\sigma_{\alpha\beta}}\sim
g_{\alpha}\dot{x}_\beta-g_\beta\dot{x}_\alpha$. Thus, for the
classical solution $g[x^{\rm cl}]$ it gives zero contribution. It
is, furthermore, easy to verify that the third term in (107) also
disappears for $x=x^{\rm cl}$. We, therefore, conclude that
\begin{equation}
\frac{\delta^2A[C]}{\delta x_\alpha^{\rm cl}(\tau)\delta
x_\beta^{\rm cl}(\tilde{\tau})}=\dot{x}_\mu^{\rm
cl}(\tau)\dot{x}_\nu^{\rm
cl}(\tilde{\tau})\Delta^{(2)}_{\mu\alpha\nu\beta}[ x^{\rm
cl}(\tau)-x(\tilde{\tau})].
\end{equation}

Inserting Eq.(107) into Eq.(108) and taking into account that the
dominant contribution to the two-point correlator comes from the
region $\tau\approx\tilde{\tau}$ we find
\begin{equation}
S\approx S_{\rm{cl}}+\int\limits_0^b
d\sigma\,y_\alpha(\sigma)\left[-{1\over2}\frac{m_0}{\mid\dot{x}\mid}\delta_{\alpha\beta}\frac{\partial^2}{\partial
\sigma^2}+\frac{\lambda^2}{T_g}\omega_{\alpha\beta}(\sigma)\right]y_\beta(\sigma).
\end{equation}
Let it be remarked that to arrive at the above relation we have
adopted the expansion of the two-point correlator indicated in Eq.
(28). We have also used the helicoid parametrization observing, at
the same time, that the eikonal lines give null contribution. One
further realizes that the contributions of the two helical curves
to the linear term in (106) cancel each other, since
$\ddot{x}^{(2)}_\mu(s)=-\ddot{x}^{(4)}_\mu(s)$ and $\tau_2-\tau_1
\simeq \tau_4-\tau_3\sim\frac{b\mid\dot{x}\mid}{2m_0}$.

The non-trivial contribution of the helical curves is incorporated
in the term
\begin{equation}
\omega_{\alpha\beta}=\delta_{\alpha\beta}-{1\over2\dot{x}^2}
\left(\dot{x}_\alpha^{(2)}\dot{x}_\beta^{(2)}+\dot{x}_\alpha^{(4)}\dot{x}_\beta^{(4)}\right),
\end{equation}
the origin of which is the second functional derivative, {\it cf.}
(106). The mass parameter $\lambda^2$ in (109) has the same source
and is defined as
\begin{equation}
\lambda^2\equiv \mid \dot{x}\mid T_g^2\int\limits _0^\infty
dw\left(D(w^2)+D_1(w^2)+{d\over dw^2}D_1(w^2)\right).
\end{equation}

The differential operator entering Eq.(109) has no zero
eigenvalues since the $``$classical'' solution is, in fact, the
one that annihilates the $g$-function. Accordingly, the
calculation of the path integral over $y=x-x^{{\rm cl}}$ does not
require any particular regularization. A straightforward
calculation shows that
\begin{equation}
\det\omega_{\alpha\beta}={1\over\dot{x}^2}\left(1-{1\over\dot{x}^2}\right)
=\frac{T^2\theta^2/b^2}{1+\theta^2/b^2}.
\end{equation}
Thus the matrix $\omega_{\alpha\beta}$ can be diagonalized and the
$y$-integral can be easily performed. However, in the limit
$m_0\to 0$ it can be immediately seen that the integration over
the boundary fluctuations gives prefactors which are powers of the
logarithm of the incoming energy and as far as Regge behavior is
concerned, they cannot change the behavior that was determined in
the previous section.

The next task is to take up the issue of the spin-field dynamics
contribution to the scattering amplitude. As seen in Section 3  a
spin factor is associated with each segment of the worldline path.
This factor receives contributions from two sources. The first one
is
\begin{equation}
\int d\tau\int\limits_{S(C)} dS\cdot\Delta^{(2)}(z-x)\cdot J=\int
d\tau\frac{\dot{x}_\mu g_\nu-\dot{x}_\nu
g_\mu}{\dot{x}^2}{i\over4}\left[\gamma_\mu,\gamma_\nu\right]
\end{equation}
and is obviously zero for the classical trajectory (72).

The other term has the form
\begin{equation}
P={1\over8}\int d\tau\int
d\tau'J_{\mu\nu}\Delta^{(2)}_{\mu\nu,\lambda\rho}(x-x')J_{\lambda\rho}={3\over4}\int
d\tau\int d\tau'(D+D_1)+{3\over8}\int d\tau\int
d\tau'(x-x')^2D'_1.
\end{equation}

In the stochastic limit, within which we are working, the
integrals in the above equation give appreciable contribution only
for $\mid
x(\tau)-x(\tau')\mid\approx\mid\dot{x}\mid\,\mid\tau-\tau'\mid\ll
T_g$. More concretely, consider the contribution to (114) from the
helical curve $(\bar{1}\to 2)$. A straightforward calculation
shows that the analytically continued result is
\begin{equation}
P=-(t_2-t_1)^2{M^4\over\chi},
\end{equation}
where we have written $\tau=it$ for the time variable and denoted
\begin{equation}
M^4=\left(8\frac{\int\limits_0^\infty
dwD(w^2)}{\int\limits_0^\infty dw\,w^2D_1(w^2)}\right)^{1/2}
\int\limits_0^\infty
dw\left(D(w^2)+D_1(w^2)+{1\over2}\frac{d}{dw^2}D_1(w^2)\right).
\end{equation}

As has been mentioned in Section 3 and discussed in [27],
contribution (115) has an interesting role as far as the form of
the fermionic propagator is concerned. As it has been shown in
[26,32] this non-perturbative $``$paramagnetic" contribution
corrects the self-energy of a bound light quark and consequently
the Regge intercept without changing the slope. We shall not
discuss here this interesting issue  leaving it for a forthcoming
study. In this paper we bypass the problem using the
parametrization [17,36].

The remaining spin structure is summarized in the chain
\begin{equation}
I=\prod\limits_{i=4}^1m_i\left[1-{1\over2m_i}\gamma\cdot\dot{x}^{(i)}(\tau_i)\right],
\end{equation}
which must be sandwiched between the external spinor wavefunctions
representing the incoming and outgoing quarks (in the simple
picture wherein the meson wavefunction is just the product of free
spinors). The non-trivial dynamics of the process are now
incorporated into the fact that the vectors
$x_\mu^{(i)},\,i=1,2,3,4$ forming the boundary of the helicoids,
are 3-dimensional vectors with $\mid\dot{x}^{(i)}\mid^2=const.$
For $i=1,3$ turns the factor in (117) to the operator
$1-\frac{\gamma\cdot p^{(i)}}{\mid p^{(i)}\mid}$.

For $i=2,4$ the matrices
\begin{equation}
I_2=1-{1\over2m}\frac{b}{\tau_2-\tau_1}\gamma\cdot\dot{x}^{(2)}(b)\to1-\frac{\gamma\cdot\dot{x}^{(2)}(b)}{\mid
\dot{x}^{(2)}\mid}
\end{equation}
and
\begin{equation}
I_4=1-{1\over2m}\frac{b}{\tau_4-\tau_3}\gamma\cdot\dot{x}^{(4)}(b)\to1-\frac{\gamma\cdot\dot{x}^{(4)}(b)}{\mid
\dot{x}^{(4)}\mid},
\end{equation}
are also representations of projection operators. As shown in [31]
the matrices (118) and (119) are the direct product of two
$2\times 2$ matrices each of which are by themselves projection
operators. Given these observations it becomes a matter of simple
algebra to find that the standard kinematics are reproduced.

\vspace*{.4cm}

  {\bf 6. Concluding Remarks}

\vspace*{.4cm}

  The central objective of this paper was to assess the merits of
the Stochastic Vacuum Model of Dosch and Simonov as a credible
representative of QCD. From a methodological standpoint we
employed the path-integral approach for the  casting of the
theory, a practice that has been proved an ideal tool for the
exploration of its non-perturbative aspects on which the present
study is focused. In the first part of the paper-and at a purely
theoretical level-we verified both the loop equations and the
Bianchi identity through the SVM, an occurrence which further
solidifies the credibility of the model. We have also derived an
explicit expression for the spin factor that represents the
non-perturbative spin-field dynamics and necessarily enters the
analysis of physical processes. In the second part of the paper we
assessed  the effectiveness of the SVM, always in its
path-integral casting, to confront a dynamical problem where a
non-perturbative treatment is essentially important. More
explicitly, we appropriately modelled a meson-meson scattering
process in the Regge kinematical regime. In a $``$semiclassical"
approximation and always working in the framework of the
Stochastic Vacuum Model we found a Regge-type behavior for the
scattering amplitude with linear Regge trajectories. The specific
methodology we followed is entirely based on the capability of the
SVM to represent the non-perturbative content of QCD and, perhaps,
it traces a way for analytically calculating Regge trajectories in
the physical region of scattering, {\it i.e.}, square momentum
transfer $t<0$.

\newpage

\vspace*{0.5cm}

\appendix
\setcounter{section}{0} \addtocounter{section}{1}
\section*{Appendix A}
\setcounter{equation}{0}
\renewcommand{\theequation}{\thesection.\arabic{equation}}
 We give here the proof of relation (20) which appears in the
 text and whose role is significant for the derivation of
 the equation that determines the surface on which the two-point
 connector {``}lives". We begin by writing the expression for
 the connector in Eq. (8):
 \begin{equation}
 \phi(z,x_0)=P\exp\left(i\int\limits _{x_0}^z dw\cdot {\cal A}\right)=P\exp\left[i\int\limits
 _0^1 d\tau\dot{w}(\tau)\cdot {\cal A}(w(\tau))\right],
 \end{equation}
 where $w_\mu(0)=x_{0\mu},\,\,w_\mu(1)=z_\mu.$
 Taking the functional derivative of (A.1) we find
 \begin{equation}
 \frac{\delta\phi}{\delta
 w_\mu(\tau')}=i\int_0^1d\tau P\exp\left(i\int_\tau^1d\tau\dot{w}
\cdot {\cal A}
(w)\right)[\partial_\tau\delta(\tau-\tau')+\delta(\tau-\tau')\dot{w}_\nu\partial_\mu
{\cal A}_\nu(w)]P\exp\left(i\int_0^\tau d\tau\dot{w} \cdot {\cal
A}(w)\right)
 \end{equation}
 or
 \begin{eqnarray}
 \frac{\delta\phi}{\delta
 w_\mu(\tau')}&=&i\delta(1-\tau'){\cal A}_\mu(z)P\exp\left[i\int\limits
 _{x_0}^z dw\cdot {\cal A}(w)\right]-i\delta(\tau')
 P\exp\left[i\int\limits
 _{x_0}^z dw\cdot {\cal A}(w)\right]{\cal A}_\mu(x_0)+\nonumber\\
 \quad&+&ig\dot{w}_\nu(\tau')P\exp\left(i\int_{w'}^z dw
\cdot {\cal
A}(w)\right)F_{\mu\nu}(w(\tau'))P\exp\left[i\int\limits
 _{x_0}^z dw\cdot {\cal A}(w)\right].
\end{eqnarray}
It, accordingly, follows that the variation of the connector reads
\begin{eqnarray}
\delta\phi&=& {\cal A}_\mu(z)\delta
z_\mu\phi(z,x_0)-i\phi(z,x_0){\cal A}_\mu(x_0)\delta x_{0_\mu}+\nonumber\\
&+&\int_0^1d\tau
\dot{w}_\nu(\tau)\phi(z,w(\tau))F_{\mu\nu}(w(\tau))\phi(w(\tau),x_0)\delta
w_\mu(\tau).
\end{eqnarray}
Keeping everything but the end point constant one immediately
deduces that
\begin{equation}
\frac{\partial\phi}{\partial
 z_\mu}=A_\mu(z)\phi(z,x_0)-\int_0^1d\tau\dot{w}_\kappa(\tau)\phi(z,w(\tau))
 F_{\kappa\lambda}(w(\tau))\phi(w(\tau),x_0)\frac{\partial w_\lambda}{\partial
 z_\mu}.
 \end{equation}

\newpage

\vspace*{0.5cm}

\appendix
\setcounter{section}{0} \addtocounter{section}{2}
\section*{Appendix B}
\setcounter{equation}{0}
\renewcommand{\theequation}{\thesection.\arabic{equation}}

In this Appendix we  present a parametrization of the functions
 $D$ and $D_1$ used
 extensively in the present paper. This parametrization
  is supported by lattice data and is extensively discussed
 in Refs. [17,36].

 The exact relations defining the functions are
 \begin{equation}
  D=\frac{\pi^2(N^2_C-1)}{2N_C}\frac{G_2}{24}\kappa D_N,\quad
  D_1=\frac{\pi^2(N^2_C-1)}{2N_C}\frac{G_2}{24}(1-\kappa)D_{1,N},
 \end{equation}
 where $D_N$ and $D_{1,N}$ are functions which determine the
 structure of the two-point correlators, as defined in [17].
 The factor $G_2$ is defined as follows
 \begin{equation}
 G_2\equiv
 \langle0\mid\frac{g^2}{4\pi^2}F_{\mu\nu}^\alpha(0)F_{\mu\nu}^\alpha(0)\mid0\rangle
 =\frac{2N_C}{4\pi^4}\Delta^{(2)}_{\mu\nu,\mu\nu}(0).
\end{equation}
For the above correlator we shall adopt the value given in Ref
[8], namely $G_2=(0.496)^4GeV^4$. The value of the numerical
quantity $\kappa$ in (B.1) is estimated in the same reference to
be 0.74. The ansatz for the function $D_N$ is [17]
\begin{equation}
D_N(z)={27\over64}\,{1\over a^2}\int d^4k e^{ik\cdot
z}\frac{k^2}{\left[k^2+\left(\frac{3\pi}{8a}\right)^2\right]^4},
\end{equation}
where
\begin{equation}
a\equiv\int\limits_0^\infty dz\,D_N(z).
\end{equation}
 A simple calculation shows
that
\begin{equation}
D_N(z)=wK_1(w)-{1\over4}w^2K_0(w), \quad w=\frac{3\pi}{8a} \mid
z\mid,
\end{equation}
with $K_\nu$ denoting a Bessel function.
The correlation length $T_g$ can be deduced from Eq. (B.5):
\begin{equation}
T_g=\frac{8a}{3\pi}.
\end{equation}
The estimated value of $a$ is
\begin{equation}
a\approx 0.35\,{\rm fm}\,{\rm or} \,\,T_g\approx 0.297 \,{\rm fm}.
\end{equation}

With the help of ansatz (B.3) and using (B.7) one can determine
the string tension:
\begin{equation}
\sigma={1\over2}T^2_g\int
d^2wD(w)={1\over2}T^2_g\frac{\pi^2(N_C^2-1)}{2N_C}{G_2\over24}\kappa\int
d^2w\left[wK_1(w)-{1\over4}w^2K_0(w)\right],
\end{equation}
or
\begin{equation}
\sigma={1\over2}T^2_g
\frac{\pi^2(2N_C^2-1)}{2N_C}\frac{G_2}{24}\kappa2\pi=a^2G_2\kappa\pi{32\over81}\approx
0.175 GeV^2
\end{equation}

The anzatz for the function $D_{1,N}$is deduced from the equation
[17,36]
\begin{equation}
\left(4+z_\mu\frac{\partial}{\partial
z_\mu}\right)D_{1,N}(z)=4D_N(z)
\end{equation}
or
\begin{equation}
D_{1,N}(z)={1\over z^4}\int_0^z dw[4w^4K_1(w)-w^5K_0(w)]
\end{equation}
The coefficient of the rigidity term entering Eq. (94) can now be
calculated:
\begin{eqnarray}
{1\over\alpha_0}&=&{1\over32}T^4_g\int
d^2w\,w^2[2D_1(w)-D(w)]=\nonumber\\&&
={1\over32}T^4_g\frac{\pi^2(N_C^2-1)}{2N_C}{G_2\over24}\int
d^2w\,w^2[2(1-\kappa)D_{1,N}(w)-\kappa D_N(w)]\nonumber\\&&
={1\over32}T^4_g\frac{\pi^2(N_C^2-1)}{2N_C}\frac{G_2}{24}2(1-\kappa)32\pi\approx
0.276.
\end{eqnarray}

\begin{center}
{\bf Acknowledgement}
\end{center}

\vspace{0.3cm}

The authors wish to acknowledge financial supports through the
research program ``Pythagoras'' (grant 016) and by the General
Secretariat of Research and Technology of the University of
Athens.



\end{document}